\documentclass[11pt,amssymb,showpacs]{iopart}
\usepackage{latexsym}
\usepackage{graphicx}
\usepackage{color}

\usepackage{citesort}

\begin{document}
% \draft \twocolumn[\hsize\textwidth\columnwidth\hsize\csname
% @twocolumnfalse\endcsname

\title[Gate voltage control of the $AlO_x/SrTiO_3$ interface electrical properties]{Gate voltage control of the $AlO_x/SrTiO_3$ interface electrical properties}
\author{J. Delahaye and T. Grenet}
\address{Institut N\'eel, CNRS $\&$ Universit\'e Grenoble Alpes, F-38042 Grenoble, France}
\ead{julien.delahaye@neel.cnrs.fr}
\date{\today} %\maketitle

\begin{abstract}
Electron-beam deposition of an insulating granular aluminium or of an off-stoichiometric amorphous alumina layer on a $SrTiO_3$ surface is a simple way to get a metallic interface from insulating materials. No heating nor specific preparation of the $SrTiO_3$ surface are needed. In this paper, we investigate how the electrical properties of this interface can be tuned by the use of a back gate voltage (electrical field through the $SrTiO_3$ substrate). We demonstrate that the slow field-effect observed at room temperature can be used to tune reversibly and in a controlled way the low temperature electrical properties of the interface. In particular, important parameters of a transistor such as the amplitude of the resistance response to gate voltage changes or the existence of an ``on'' or an ``off'' state at zero gate voltage and at low temperature can be adjusted in a single sample. This method should be applicable to any $SrTiO_3$-based interface in which oxygen vacancies are involved and might provide a powerful way to study the metal or superconductor insulator transition observed in such systems.
\end{abstract}
 \bigskip
 %End of title in one column mode

\maketitle

%resisè_In this paper, we investigate how the electrical properties of this interface can be tuned by the use of a back gate voltage (electrical field through the $SrTiO_3$ substrate). Large but slow resistance changes are observed at room temperature which are reversible under gate voltages of opposite signs. At 4K, fast resistance changes which can reach six orders of magnitude dominate the gate voltage response and go along with irreversible and memory effects. The respective roles of oxygen vacancies electro-migration and standard charge injection processes are discussed. We demonstrate how the slow field-effect at room temperature can be used to tune finely and in a controlled way the low temperature electrical properties of the interface. This last method could be extended to any $SrTiO_3$-based interface in which oxygen vacancies are involved and might provide a powerful way to study the metal or superconducting insulator transition observed in such systems.

\section{Introduction}

Various techniques have been successfully used through the last 50 years to put insulating $SrTiO_3$ (STO) crystals into a metallic state. With standard chemical doping and high temperature annealing under vacuum \cite{FrederiksePR1964,SpinelliPRB10}, the metallic state extends over the bulk of the crystal. With ion-milling \cite{ReagorNatMat05,KanNatMat05,NgaiPRB10,HerranzJAP10,GrossJAP11}, UV exposure \cite{KozukaPRB07,MeevasanaNatMat11}, UHV cleaving \cite{SantanderNature11} and oxide layer deposition \cite{ChenNatCom13,CarreteroThesis10,ThielThesis09,BastelicNatMat08,SingPRL09,ReyrenScience07}, the metallic state can be confined close to the STO surface \footnote{Note that if the thickness of the metallic layer can be of less than 10nm for oxide heterostructures, it is much larger (about 100nm or even more) for ion-milled surfaces.}. The oxide layer deposition technique was first restricted to the epitaxial growth of oxides by pulsed laser deposition at high temperature, the most famous example being the $LaAlO_3/SrTiO_3$ (LAO/STO) heterostructure \cite{OhtomoNature04,ThielThesis09,HuijbenThesis06,CarreteroThesis10}. But surface metallic states were also recently observed by using pulsed laser \cite{ChenNanoLetters11,LiuPRX13,FuchsAPL14,ScigajSSI15} and e-beam \cite{DelahayeJAPD12} depositions of amorphous oxides at room temperature. The respective roles of oxides non stoichiometry (oxygen, cations), adsorbates, ions inter-diffusion and electronic reconstruction in the formation of this surface metallic state remain an active and controversial issue \cite{OhtomoNature04,NakagawaNatMat06,KalabukhovPRB07,SiemonsPRL07,HerranzPRL07,ThielThesis09,CancellieriEPL10,HerranzSR12,ChenNatCom13,LiuPRX13,BreckenfeldPRL13,AsmaraNatCom14,LiSR15,GariglioJPCM15,ScheidererPRB15}.

In this article, we report on electrical field effect measurements of STO crystals, on which an insulating granular aluminum or an oxygen deficient alumina layer (thereafter referred to as the AlOx layer) was deposited at room temperature by electron gun evaporation. We have demonstrated recently that such deposition can put the STO surface into a metallic state \cite{DelahayeJAPD12}. The simplicity of the manufacturing process (no heating, no surface preparation) makes this method very attractive compared to the more sophisticated techniques currently used. The most likely origin of this metallic state is the formation of oxygen vacancies in the STO substrate close to the AlOx/STO interface, the oxygen being ``pumped off'' from STO when the AlOx layer is deposited on top \cite{DelahayeJAPD12}. Oxygen vacancies in STO are known to release electrons for the conduction and can lead to a metallic state if their concentration is large enough. The exact thickness of this metallic state is not known but its electrical parameters (charge carrier surface density and mobility, sheet resistance value and temperature dependence, etc.) are very similar to the 2D electron gas obtained by the pulsed laser deposition of oxides, which strongly suggests that it is confined close to the interface \cite{DelahayeJAPD12}.

When STO crystals are doped in the bulk by chemical impurities or oxygen vacancies, the metal-insulator transition occurs at small charge carrier densities compared to other oxides or even standard doped semi-conductors. The 3D critical charge density which corresponds to the metal-insulator transition is not precisely known but metallic states are reported for impurity densities as low as $10^{16}cm^{-3}$\, \cite{SpinelliPRB10}. The high value of the STO dielectric constant, especially at low temperature, may explain this striking property. At 2D, a metallic-like behaviour is observed for surface charge densities as low as a few $10^{13}cm^{-2}$\ \cite{ThielScience06}. %\cite{KozukaPRB07}
Such a value corresponds to the surface charge density that can be added or removed in a standard field effect experiment, where a gate voltage is applied between the STO surface and a metallic gate over an insulating material (the gate insulator). STO is thus a system in which large modulations of the electrical resistance are expected upon the application of a gate voltage \cite{AhnRMP06}.

A large number of electrical field effect experiments have been performed on STO crystal based devices.  Many different situations were explored: the STO surface was in the ``on'' (metallic or superconducting) state or in the ``off'' (insulating) state when no gate voltage was applied, the bulk of the STO crystal was used as the gate insulator (``back gate'' geometry) or another insulating material was deposited on top (``top gate'' geometry), the temperature of the measurement was 300K or much lower (4K), etc. The observed resistance response to gate voltage changes are also quite various: they can be fast or slow, small or with relative changes of many orders of magnitude, associated with memory effects and hysteresis, etc.
\cite{YoshidaJJAP96,PallechiAPL01,UenoAPL03,InoueCERC04,ShibuyaAPL04,ShibuyaAPL06,NakamuraAPL06,ThielScience06,ShibuyaJAP07,ShibuyaAPL08,CavigliaNature08,UenoNatMat08,BellPRL09,CenScience09,ThielThesis09,NgaiPRB10,NishioJJAP10,LeePRL11,LiPRL12,ForgAPL12,ChristensenAPL13,HosodaAPL13,EyvazovSR13,EerkesAPL13,BiscarasSR14,GallagherNP14,LiuAPLM15,HurandSR15}.
%Beyond future and hypothetical applications, the electrical field effect results can bring precious information on the conduction process of the STO surface.
But in all these studies, the electrical properties of the interface are determined by the fabrication parameters. What we show in this article is that the low temperature electrical properties of our AlOx/STO interface can also be changed after its making.
% Beyond potential applications, the electrical field effect has proven to be a powerful tool in order to tune and study the electrical properties of STO-based interfaces, such as the superconductor - insulator transition observed at the LAO/STO interface \cite{CavigliaNature08}.

%En dépit de sa grande simplicité, cette méthode est encore peu utilisée actuellement. Altough not Since this metallic state is easy to obtain (no heating, no surface preparation, etc.), it is interesting to know how its response to an electric field compares with the other and more sophisticated STO-based systems.

We studied the field effect in AlOx/STO interfaces in the ``back gate'' geometry from room T to 4K.
Our main results can be summarized as follows. At room temperature, the response to the application of a non-zero gate voltage ($V_g$) is dominated by slow changes of the resistance, which can increase by three orders of magnitude the resistance of a metallic-like interface. Since this slow resistance response is reversible and practically frozen below $\simeq 250K$, it can be used to tune the electronic state (metallic or insulating) of a given starting metallic-like interface, and to stabilize a large range of low temperature properties. For example, a state showing a huge field effect at 4K could be obtained (resistance changing by 6 orders of magnitude with a gate field of 0.6kV/cm). % Memory effects and an irreversible increase of the $V_g = 0V$ resistance are also observed at low temperature during $V_g$-cycles.
The respective roles of oxygen vacancies electro-migration and standard charge injection processes will be discussed.

%  During a ˜10 hours stay under $V_g = -30V$ (electrical field of $0.6kV/cm$), the sheet resistance ($R_s$) is found to increase from $20k\Omega$ to $10M\Omega$. The dynamics of these slow resistance changes is thermally activated around room temperature with an activation energy of the order of $0.7eV$. Such a value is in agreement with the electro-migration of oxygen vacancies in STO crystals. This slow field effect can be used to tune reversibly and on a controlled way the low temperature properties of the samples, such as their $R_s$ value and $V_g$ response. With this last method, fast resistance changes of more than 6 orders of magnitude were obtained at $4K$ under an electrical field of $0.6kV/cm$. Memory effects and an irreversible increase of the $V_g = 0V$ resistance are also observed at low temperature during $V_g$-cycles, which physical origin remains unclear (ferroelectricity, trapped electrons, etc.).
%\end{enumerate}

\section{Elaboration and measurement techniques}

The samples were made according to Ref. \cite{DelahayeJAPD12}. STO crystals one side polished, (100) oriented and $0.5mm$ thick were purchased from Neyco company. The polished surface was simply cleaned by successive ultrasonic bath in trichloroethylene, acetone and alcohol before being mounted in an electron beam evaporator. Al contacts, $20nm$ thick, were deposited first. Then, $40nm$ of insulating granular Al or $5nm$ of $O_2$ deficient alumina passivated by $95nm$ of stoichiometric alumina was deposited between the Al contacts without breaking the vacuum. The $O_2$ deficient alumina layer was obtained by the evaporation of alumina at $0.5\AA /s$ under an $O_2$ pressure of less than $10^{-5}mbar$, the stoichiometric alumina layer by the evaporation of alumina at $0.5\AA /s$ under an $O_2$ pressure of $2\times 10^{-4}mbar$ and the insulating granular Al layer by the evaporation of pure Al at $1.8\AA /s$ under an $O_2$ pressure of $4-5\times 10^{-5}mbar$. The base pressure of the evaporator is less than $10^{-6}mbar$. As long as the sheet resistances of the samples are the same, we did not observe any significant difference between the field effect response of granular Al and $O_2$ deficient alumina / STO interfaces.

\begin{figure}[h]
%Schema
\centering
\includegraphics[width=7cm]{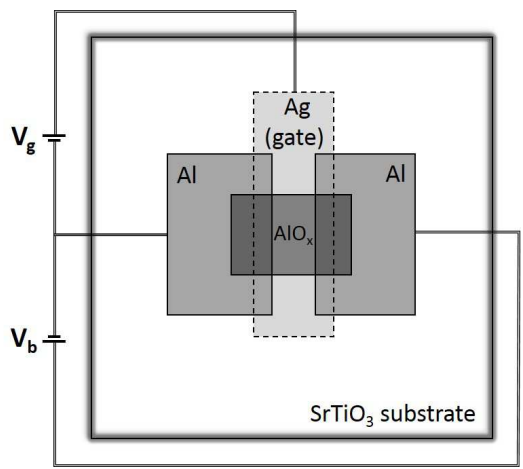}
\caption{Sketch of a typical sample: the STO substrate is $10 \times 10 \times 0.5 mm$, the Al contacts $4 \times 2.5mm$, the active AlOx channel (the part between the Al contacts) $\simeq 2 \times 1mm$ and the gate width (on the other side of the STO substrate) is $\simeq 2mm$. The polarities of the gate ($V_g$) and bias ($V_b$) voltages are also indicated.}\label{Figure2}
\end{figure}

The active channel between the Al contacts (i.e. the part of the surface covered by the AlOx overlayer) has a typical size of $1mm \times 2mm$. Its electrical resistance was measured in a two contact MOSFET-like configuration \footnote{For metallic samples, two and four contacts configurations could be used and no significant difference was observed.} and unless otherwise specified, in the ohmic regime (linear part of the $I-V_b$ curves, $V_b$ being the bias voltage). Depending on the resistance value, either current bias or voltage bias were used. In our electrical field effect measurements, the gate insulator is the STO substrate itself and no leakage currents were detected with the $V_g$ values used (leakage currents below $1pA$, maximum absolute $V_g$ value of $100V$ corresponding to an electric field of $2kV/cm$). The bias voltage $V_b$ was usually kept much smaller than $V_g$ in order to avoid any mix up between the two parameters. The gate contact was made on the unpolished side of STO crystal with silver paint and its polarity is such that it is connected to the plus terminal of the $V_g$ source (see Figure \ref{Figure2}).

\section{Room temperature modulation of the resistance}

\begin{figure}[h]
% RTSTO
\centering
\includegraphics[width=8cm]{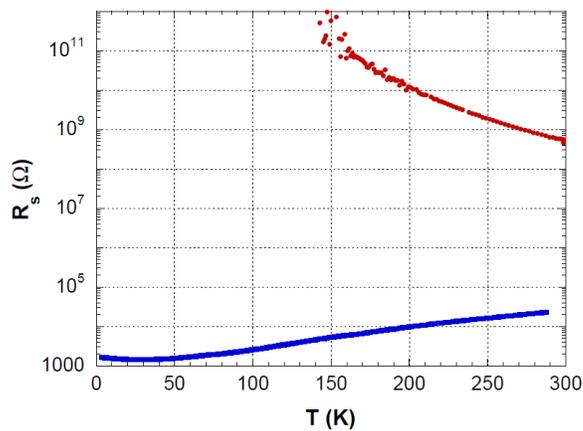}
\caption{$R_s$ versus T between $4K$ and $300K$ for two ``extreme'' AlOx/STO samples. The high-$R_s$ sample was kept 3 days in the dark before the measurement. The highest resistance measurable in our experimental set-up is of $10^{10} - 10^{11} \Omega$.}\label{Figure1}
\end{figure}

Depending on the oxygen pressure (Al/O ratio during the evaporation), we can get samples with room temperature $R_s$ values from $\simeq 20k\Omega$ to unmeasurably large values \cite{DelahayeJAPD12}. Typical $R_s-T$ curves of low and high-$R_s$ samples are plotted on Figure \ref{Figure1} in the range $4K - 300K$. The resistance of a $R_{s300K} = 30k\Omega$ sample decreases by a factor of 10 between $300K$ and $30K$, with a small resistance increase at lower temperature as observed elsewhere for samples with similar $R_s$ values \cite{BellPRL09,HerranzPRL07}. The resistance of a $R_{s300K} = 500M\Omega$ sample displays instead a fast increase (close to an exponential) when the temperature is lowered and is already not measurable (R above $100G\Omega$) around $150K$. The transition from a room temperature metallic-like (temperature coefficient $dR/dT$ positive around $300K$ \footnote{Such samples are not metallic in the strict sense since most of them display a clear diverging resistance at low temperature.}) to a room temperature insulating-like ($dR/dT$ negative around $300K$) behaviour occurs around $R_{s300K}\simeq 1M\Omega$. We will focus thereafter on metallic-like interfaces, having $R_s$ values in the range $20-30k\Omega$ at 300K. % Such samples are pretty stable over time and the resistance drifts are limited to $20\%$ six months after their making.} \textcolor[rgb]{1.00,0.00,0.00}{% The stability in time of such samples, the resistance drift observed after their making stays below $\simeq 20\%$ after six months of measurements.
A resistance increase of $\simeq 10\%$ which saturates within about 2 hours is observed when the samples are transferred from ambient daylight to darkness, and all the following measurements were thus performed after at least one day in the dark.

\begin{figure}[h]
% GVg300KLowRs et GVg300KLowRsZoom2
\centering
\includegraphics[width=8cm]{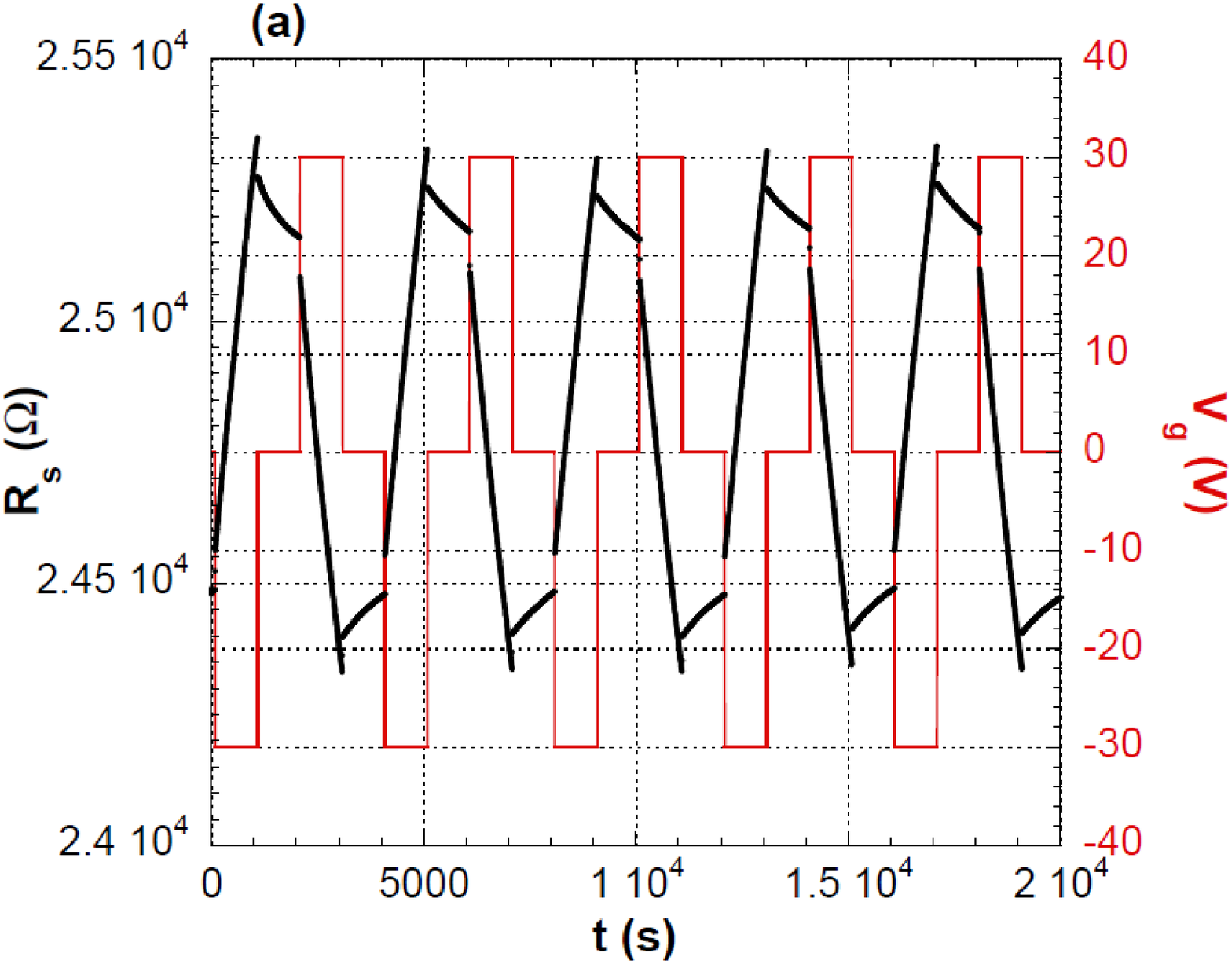}
\includegraphics[width=8cm]{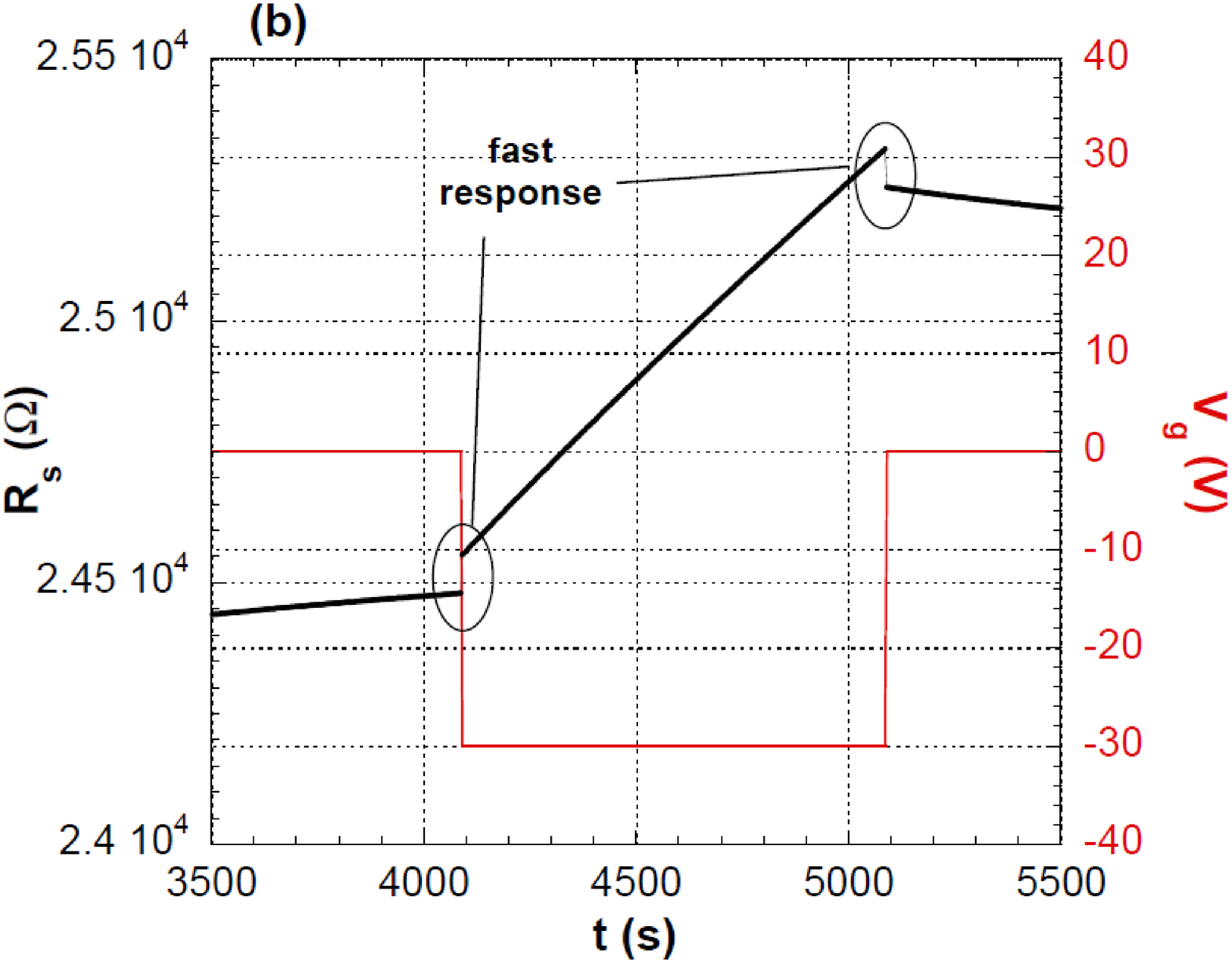}
\caption{(a) $R_s$ as a function of time under repeated $V_g$ cycles $0V, -30V, 0V, 30V$ at room temperature. (b) Zoom of the top figure over a reduced time scale. The slow $R_s$ drift is about 3\% over $1000s$ while the fast shifts corresponding to a $V_g$ change of $\pm 30V$ are equal to 0.33\%.}\label{Figure3}
\end{figure}

Typical behaviours of such low $R_s$ samples submitted to repeated gate voltage cycles ($V_g = 0V, +30V, 0V, -30V$) are plotted in Figure \ref{Figure3}. Such $V_g$ cycles are commonly used in STO-based field effect measurements in order to quantify fast and slow responses to $V_g$ changes and to reveal a potential memory of the $V_g$ values experienced by the sample. Apart from small fast (faster than $\simeq 1s$) $R_s$ changes occurring when $V_g$ is switched, the $R_s$ response is found to be dominated by slow variations. Each stay $\simeq 1000s$ long under a positive $V_g$ results in a slow resistance decrease with the opposite trend under a negative $V_g$. Moreover, during the $0V$ stay, the resistance keeps a memory of the latest $V_g$ experienced by the sample: when $V_g$ is switched back to $0V$, the resistance tends to come back to its value before the latest $V_g$ change. But these drifts are much slower than under positive or negative $V_g$ and the $R_s$ restoration is only partial. The $R_s$ values observed over one $V_g$ cycle are therefore not symmetrical to the initial $0V$ value. $R_s$ modulations are reproducible over many gate voltage cycles and typical values of slow and fast resistance variations are indicated in the legend of Figure \ref{Figure3}. Interestingly enough, the resistance modulations are strongly reduced when the samples are exposed to daylight (see Supplementary materials A).

\begin{figure}[h]
  % GVg300KLowRsLongt
  \centering
\includegraphics[width=8cm]{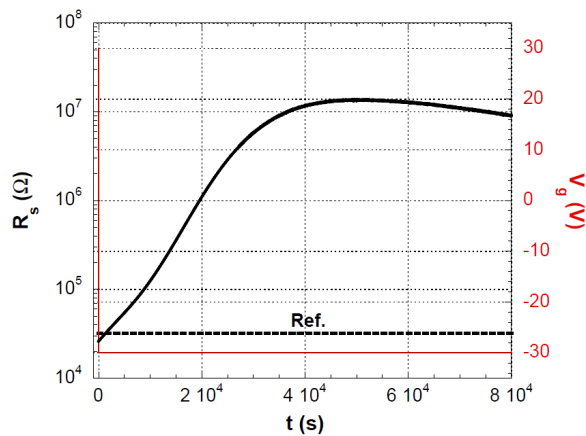}
\caption{$R_s$ response following a $V_g$ change from $30V$ to $-30V$ after the samples have been allowed to reach their minimum resistance values under $V_g = 30V$. The $R_s$ values before any $V_g$ change are indicated by the dotted lines.}\label{Figure4}
\end{figure}

  Much larger $R_s$ changes are obtained when non-zero $V_g$ are applied over longer times. Starting from a ``fresh'' (no $V_g$ history) low-$R_s$ sample ($R_s = R_{sref}=30k\Omega$), $R_s$ reaches a minimum value about $10\%$ smaller after few hours under $V_g = 30V$, while a steady $R_s$ increase is observed over $20h$ under $V_g = -30V$, up to a maximum value of $\simeq 10M\Omega$ (see Figure \ref{Figure4}). Such a value is already beyond the metallic-like to insulating crossover of $\simeq 1M\Omega$ discussed before.
  % Comparatively, $R_s$ is reduced by only 10\% compared to $R_{sref}$ when a gate voltage of $+30V$ is applied.

  % In high-$R_s$ samples, the effects are qualitatively similar but the relative changes are larger for both $V_g$ signs. Starting from a ``fresh'' sample having $R_s = R_{sref} = 800M\Omega$, $R_s$ drops to $20M\Omega$ under a long stay under $V_g = 30V$ (minimum resistance value), and becomes unmeasurably large ($> 10^{11}\Omega$) after ˜$3h$ under $V_g = -30V$ (see Figure \ref{Figure4}). In low-$R_s$ samples, the resistance can be tuned reversibly many times between its maximum and minimum values simply by using long $V_g$ stays of opposite signs.

 The fast field-effect can be simply understood as a standard $V_g$ induced charge injection or removal in the system. The fact that a $V_g$ increase is associated with a resistance drop, and a $V_g$ decrease with a resistance jump is in qualitative agreement with the negative sign of the charge carriers obtained by Hall effect measurements \cite{DelahayeJAPD12}. We can go one step further and use the amplitude of the resistance jumps or drops to estimate the surface charge carrier density $n_s$ at the interface. If all the charge carriers involved in the conduction have the same mobility \footnote{The charge carrier mobilities were found to be the in the range $1-10cm^2.V^{-1}.s^{-1}$ for surface and bulk metallic-like states \cite{HerranzPRL07,SpinelliPRB10}. This approximation should thus be reasonable at least for low-$R_s$ samples.}, the relative amplitudes of resistance jumps or drops $|\Delta R_s/R_s|$ should be equal to the relative changes in the surface charge carrier density $|\Delta n_s/n_s|$ (as long as the relative changes are smaller than 1). Assuming a simple plane-plane capacitance geometry, $\Delta n_s(\Delta V_g) = (\epsilon/d)\Delta V_g$ where $\epsilon$ is the dielectric constant of the STO substrate ($300\epsilon_0$ at $300K$) and $d$ its thickness ($0.5mm$).
 For the sample of Figure \ref{Figure3} ($R_s = 25k\Omega$), $\Delta R_s(30V)/R_s$ and thus $\Delta n_s(V_g)/n_s$ are equal to $0.3\%$. A $V_g$ change of $30V$ corresponds at $300K$ to $\Delta n_s(30V) = 1.0\times 10^{11}e/cm^2$, which gives a $n_s$ estimate of  $3\times 10^{13}e/cm^2$. This $n_s$ value is in quantitative agreement with Hall effect measurements on samples of similar $R_s$ \cite{DelahayeJAPD12}.
 %The same analysis on the high-$R_s$ sample of Figure \ref{Figure3} ($R_s = 800M\Omega$) gives $\Delta R_s(30V)/R_s = \Delta n_s(30V)/n_s = 10\%$ and thus $n_s = 1.0\times 10^{12}e/cm^2$. Comparing these two samples, $R_s$ values differ by a factor of 3000 while the $n_s$ values differ by only a factor of 30, which strongly suggests that the large $R_s$ difference is mainly due to a difference in the charge carrier mobilities.

Let's now discuss the prominent slow part of the field effect. In our samples, the charge carriers are supposed to be electrons released by oxygen vacancies in the STO substrate. When $V_g\ne 0$, an electrical field exists in the bulk of the STO substrate up to the AlOx/STO interface conducting state. Note that this external field adds to a possible internal field present at the interface when $V_g= 0$. According to different experiments, oxygen vacancies have a significant mobility in STO at room temperature \cite{GrossJAP11,SchultzAPL07,ChristensenAPL13}. It is thus tempting to explain the slow $R_s$ changes observed under non-zero $V_g$ as slow drifts of the oxygen vacancies under this electric field. If $V_g < 0$ (resp. $> 0$), the electrical field is such that it pulls (resp. pushes) the positively charged oxygen vacancies further from (resp. closer to) the interface. The concentration of oxygen vacancies close to the AlOx/STO interface thus decreases (resp. increases) under negative (resp. positive) $V_g$. By analogy with what occurs in disordered induced metal-insulator transition, the decrease of the charge carrier concentration is accompanied by a strong suppression of their mobility below some critical concentration \cite{GrossThesis09}.
% The large $R_s$ changes observed in high-$R_s$ samples under both $V_g$ polarities and in low-$R_s$ samples under negative $V_g$ may simply reflect the fact that the carrier concentration lies below the critical one.
The crucial role of mobility changes is supported by a quantitative analysis on the sample of figure \ref{Figure4}: its surface charge carrier density (deduced from fast resistance jumps and drops, see before) is divided by only 7 between the minimum and maximum $R_s$ states while $R_s$ is multiplied by 400 \footnote{The fact that the resistance goes through a maximum under negative $V_g$ in low-$R_s$ samples cannot be understood within this simplified picture. It might result from the combined effects of an internal electrical field at the AlOx/STO interface and of a charge mobility dependence with the distance to the interface.}. When $V_g$ is switched back to $0V$, the $V_g$ induced electro-migration stops and the resistance shifts are strongly reduced, giving rise to the memory effect described before.

Slow resistance drifts and memory effects are indeed quite common at room temperature in STO-based field effect devices \cite{ThielThesis09,ThielScience06,ChristensenAPL13}. They are usually attributed to the electro-migration of oxygen vacancies in the STO substrate, an hypothesis which is strengthened by the fact that these features are smaller or absent when top gated insulators are used
\cite{HosodaAPL13,ThielScience06,EyvazovSR13,ForgAPL12}. But our explanation based on a strong concentration dependence of the charge carriers mobility doesn't seem to be universal: in amorphous LAO/STO heterostructures, the large and slow $R_s$ changes are accompanied by large changes of the surface charge density, with only a negligible alteration of the mobility \cite{ChristensenAPL13}.

\begin{figure}[h]
% GtT300KLowRs et GrT300KLowRsScale
\centering
\includegraphics[width=8cm]{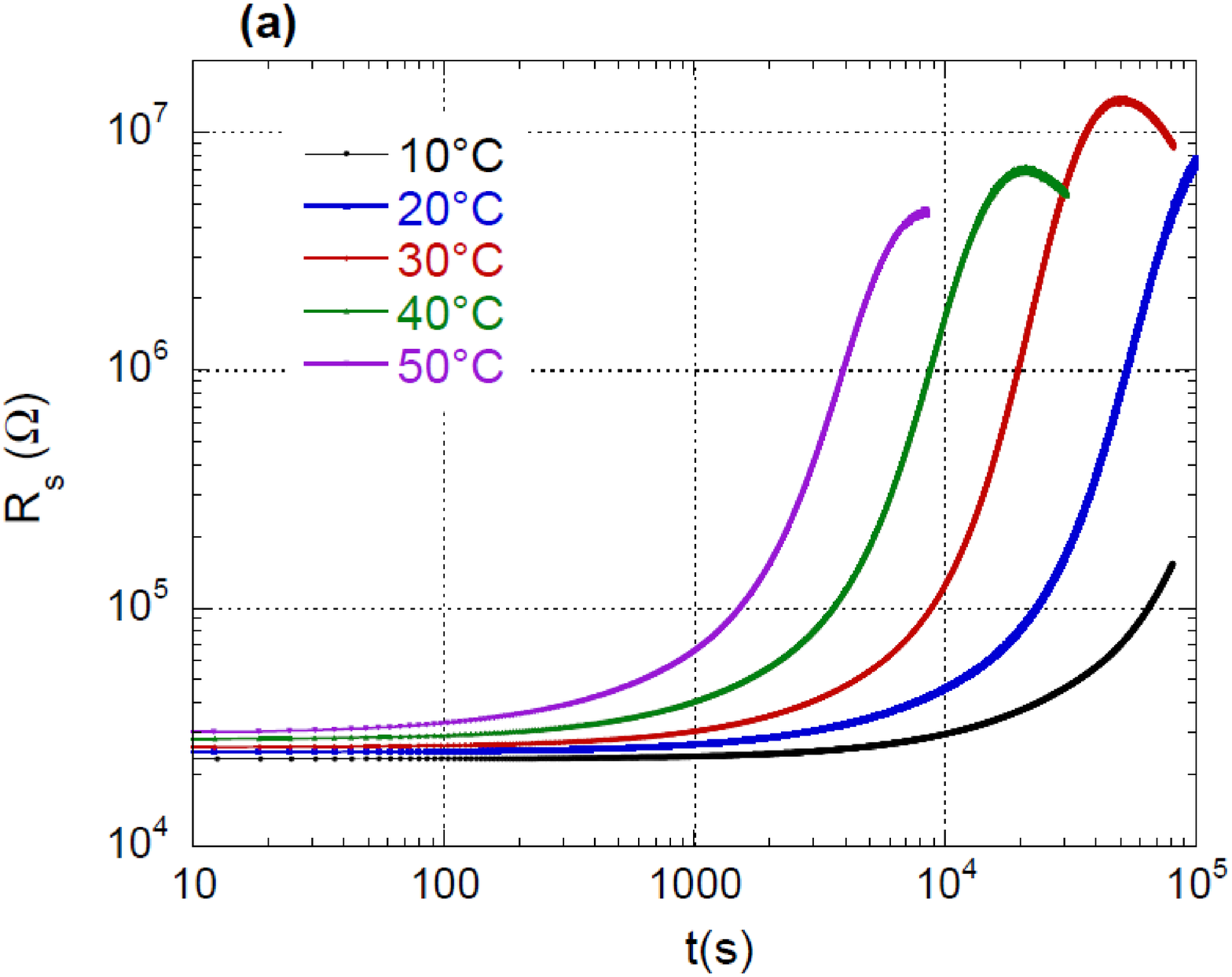}
\includegraphics[width=8cm]{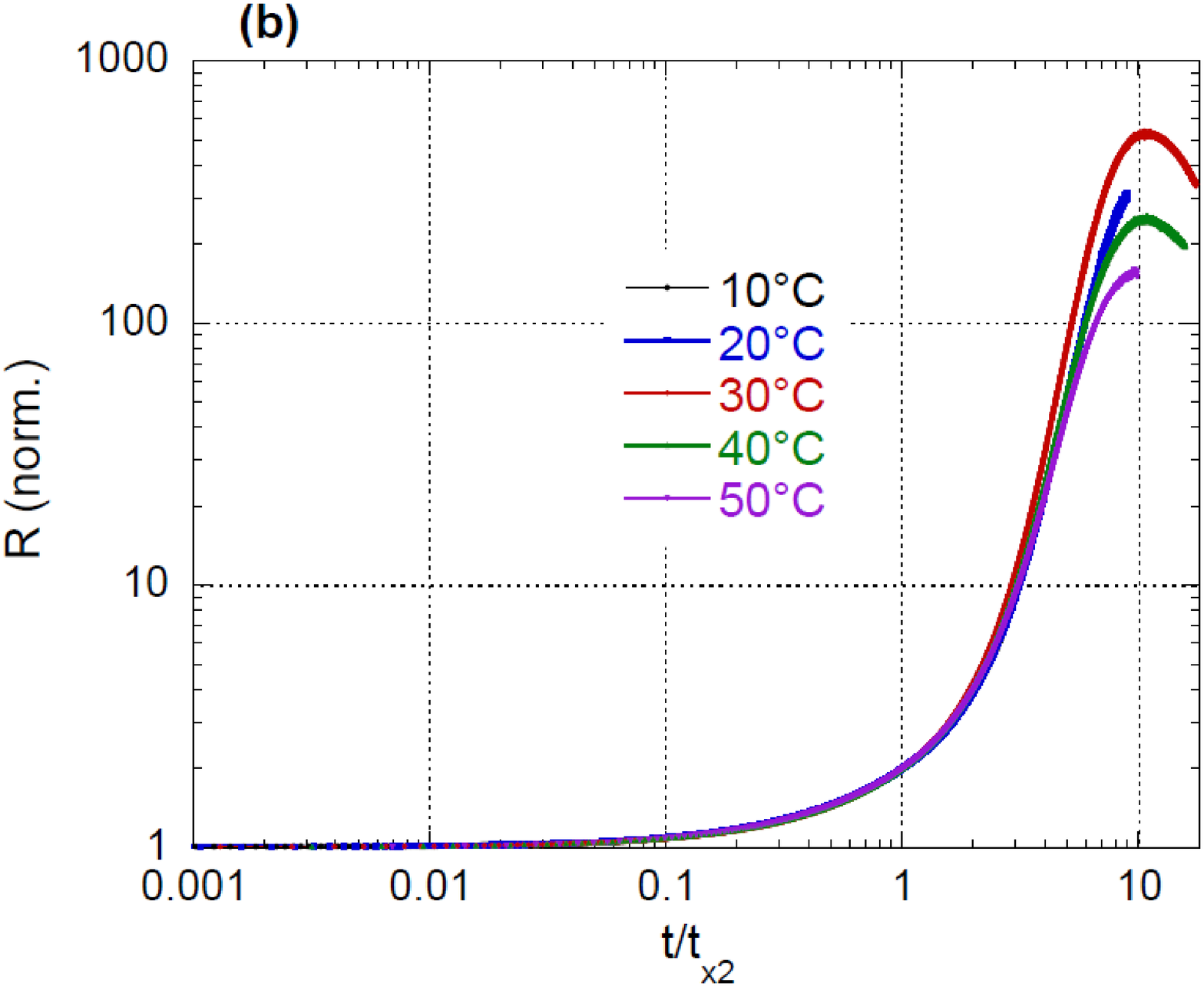}
\caption{(a) $R_s$ response to a $V_g$ change from $30V$ to $-30V$ and for different T between $10^oC$ and $50^oC$ (see the text for details). (b) $R_s$ normalized to the 30V reference value plotted as a function of $t/t_{\times 2}$ where $t_{\times 2}$ is the time at which $R_s$ has been multiplied by 2 at a given T.}\label{Figure5}
\end{figure}

\begin{figure}[h]
% GtT300KTimes2
\centering
\includegraphics[width=8cm]{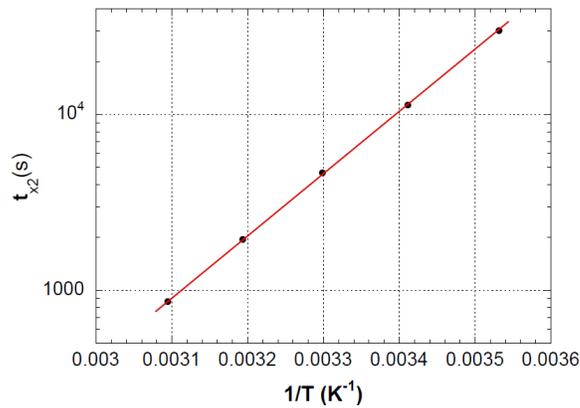}
\caption{Characteristic times $t_{\times 2}$ extracted from the scaling analysis of $R_s(t,V_g, T)$ data (see Figure \ref{Figure5}). The straight line corresponds to an activated behaviour $t_{\times 2} = A \exp(-T_0/T)$, with $T_0 = 8100K$.}\label{Figure6}
\end{figure}

In order to test in more detail the oxygen vacancy electro-migration hypothesis, we have studied how the slow $R_s$ drifts were affected by temperature changes around $300K$. Our protocol was the following. The sample was first let to equilibrate under $30V$ until $R_s$ reached its minimum value. Then, $V_g$ was changed to $-30V$ and the subsequent $R_s$ increase was measured as a function of time. The same protocol was repeated at different temperatures between $10^oC$ and $50^oC$. Typical results are plotted in Figure \ref{Figure5}. It is clearly seen that the $R_s$ changes are strongly slowed down when T is reduced. Moreover, all the curves can be merged together far from the saturation regime by normalizing $R_s$ to its value at $30V$ and the time scale by an ad-hoc characteristic time for each T (see the legend of Figure \ref{Figure6} for the exact definition of this characteristic time). Even if the T range is small in kelvin scale, the T dependence of the characteristic times is close to an activated behavior (see Figure \ref{Figure6}), with an activation energy of ˜$0.7eV$ \footnote{The exact definition of the characteristic times will change the time scale in Figure \ref{Figure6} but not the activation energy extracted.}. A similar value was found in a much more resistive sample, having $R_{s300K}=800M\Omega$. This value of $0.7eV$ is in agreement with previous experimental and theoretical estimates for oxygen vacancy diffusion coefficient in STO \cite{CorderoPRB07,SchultzAPL07,CuongPRL07}, which strengthens our hypothesis that the electro-migration of oxygen vacancies (isolated or as clusters) is responsible of the slow resistance drifts observed under $V_g \neq 0$.  In amorphous LAO/STO interfaces \cite{ChristensenAPL13}, no quantitative estimate of the activation energy was done, but the resistance drifts are absent below $\simeq ˜270K$, in qualitative agreement with the strong temperature dependence highlighted in Figures \ref{Figure5} and \ref{Figure6}.

\section{$V_g$ tuned interfaces measured at low temperature}

\begin{figure}[h]
% RTModVg1
\centering
\includegraphics[width=8cm]{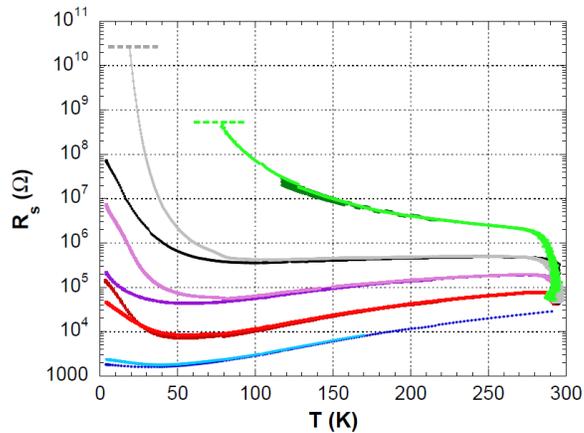}
\caption{$R_s-T$ curves measured on a low-$R_s$ sample after successive $V_g$ modulations of its resistance at $300K$ (see the text for details). Dark and light colors are used respectively for the cool-down and the warm up of the sample (after $V_g$ cycles at $4K$). All the curves were measured under $V_g=0V$ expect the dark red one which corresponds to a cool down under $-30V$. The near vertical parts of the curves observed around room temperature reflect the stays under -30V which are performed prior to the cool-down in order to adjust the $R_s$ value of the sample, and the slow recovery of the initial $R_s$ value under $0V$ when the sample is warmed up to room temperature.}\label{Figure7}
\end{figure}

We are now going to demonstrate how this strong temperature dependence of the slow field effect can be used to tune the low temperature properties of a unique AlOx/STO interface. The protocol is as follows.
%This strong T dependance of the field effect can be used to tune the low T properties of a unique AlOx/STO sample. In order to do so
Starting with a metallic-like sample having $R_{s300K} = 20-30k\Omega$, we first apply a negative $V_g$ of $-30V$ until the desired $R_s$ value is obtained (values up to $10M\Omega$ can be achieved, see before). Then, we quickly cool down the sample below $\simeq 250K$, usually under $V_g = 0V$, in order to freeze the sample parameters. At this temperature, the characteristic time of the slow resistance drifts are already so long that they cannot be measured in practice. Typical $R_s-T$ curves of such $V_g$-tuned sample are plotted in Figure \ref{Figure7}. It is seen that, like in non $V_g$-tuned samples, the $R_s-T$ behavior can indeed be changed from metallic-like ($dR_s/dT>0$ around 300K) to insulating-like ($dR_s/dT<0$ around 300K) when $R_s$ at $300K$ exceeds $˜\simeq 1M\Omega$. A large $R_s$ range from a few $k\Omega$ to non measurable resistances ($R_s > 10G\Omega$) can be obtained at 4K. The low T differences observed between cool-down and warm-up curves are due to the irreversible resistance increase which occurs during the first $V_g$ cycles at low T (see Supplementary materials B).

% We have seen in the previous section that a temperature decrease of only a few tens of degrees around room T strongly slows down the resistance response to $V_g$ changes. A natural question is thus: how the electrical field effect looks like at much lower temperatures, such as $4K$? In insulating samples ($R_{s300K}$ above $1M\Omega$), the resistance is already immeasurably large at $77K $(above $10G\Omega$) and we will limit our discussion to metallic-like samples with smaller $R_s$ values.

\begin{figure}[h]
% GVg4K50V10V
\centering
\includegraphics[width=8cm]{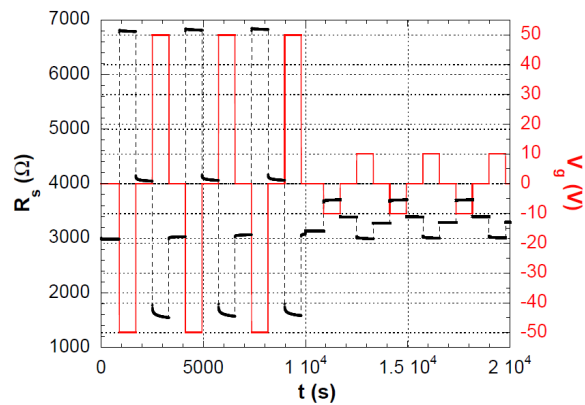}
\caption{$R_s$ versus time during $50V$ and $10V$ $V_g$ cycles at $4K$ for a low-$R_s$ sample ($R_{s300K} = 30k\Omega$).}\label{Figure9}
\end{figure}

The typical $4K$ response of a non $V_g$-tuned sample ($R_{s300K} = 30k\Omega$) to repeated $V_g$ cycles (0V, $-V_{g0}$, 0V, $V_{g0}$) is plotted on Figure \ref{Figure9}. The fast field effect now dominates the $R_s$ modulation and is much larger than at $300K$ for the same sample: $R_s$ is multiplied by more than 2 after a $V_g$ change from $0V$ to $-50V$, compared to an increase of less than 1\% at $300K$.
A memory effect is also present, i.e. the $R_s$ value at $V_g$ = 0V depends of the previous $V_g$ applied.

\begin{figure}[h]
% GVg4KMod1 et GVg4KMod2
\centering
\includegraphics[width=8cm]{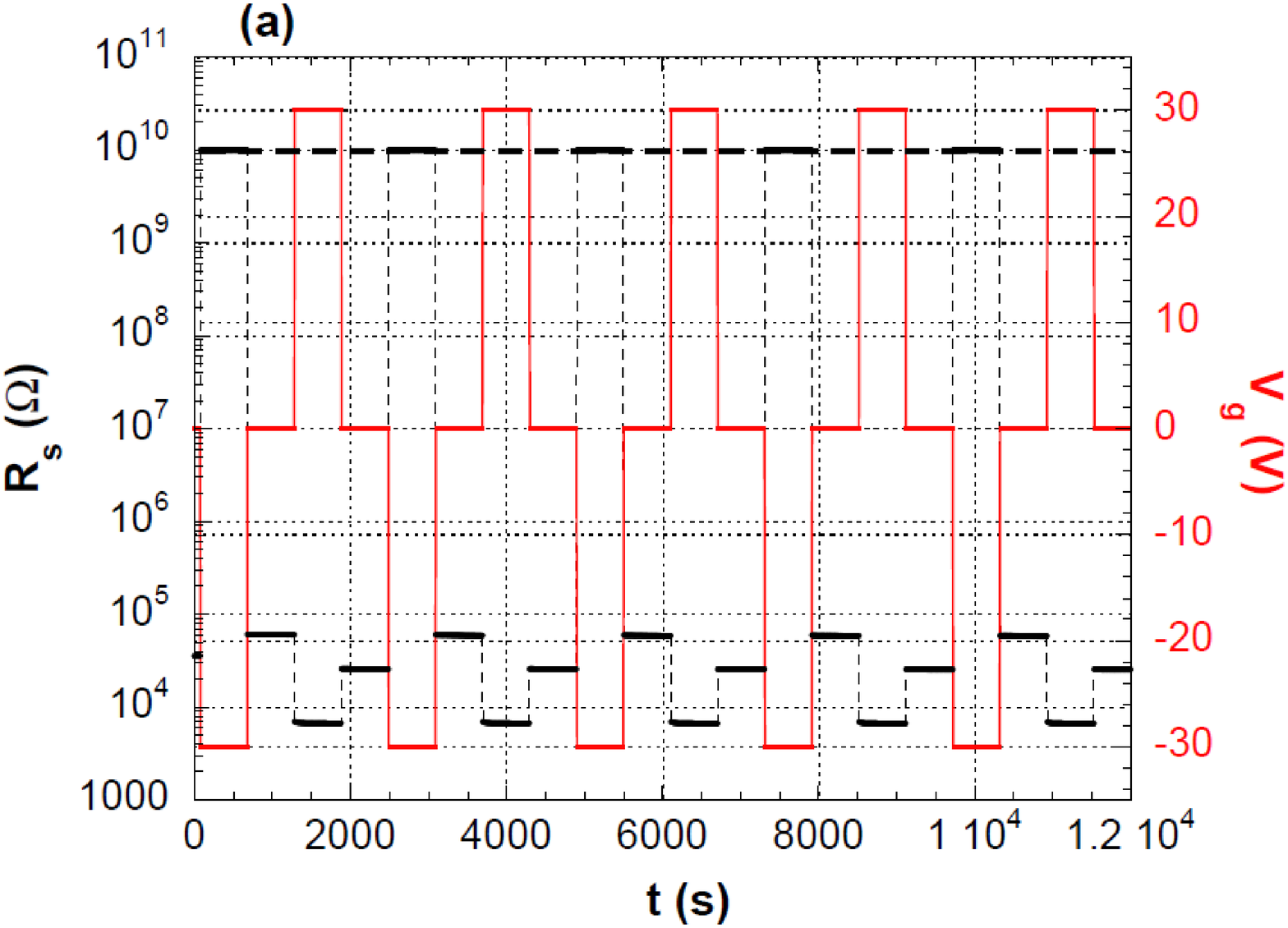}
\includegraphics[width=8cm]{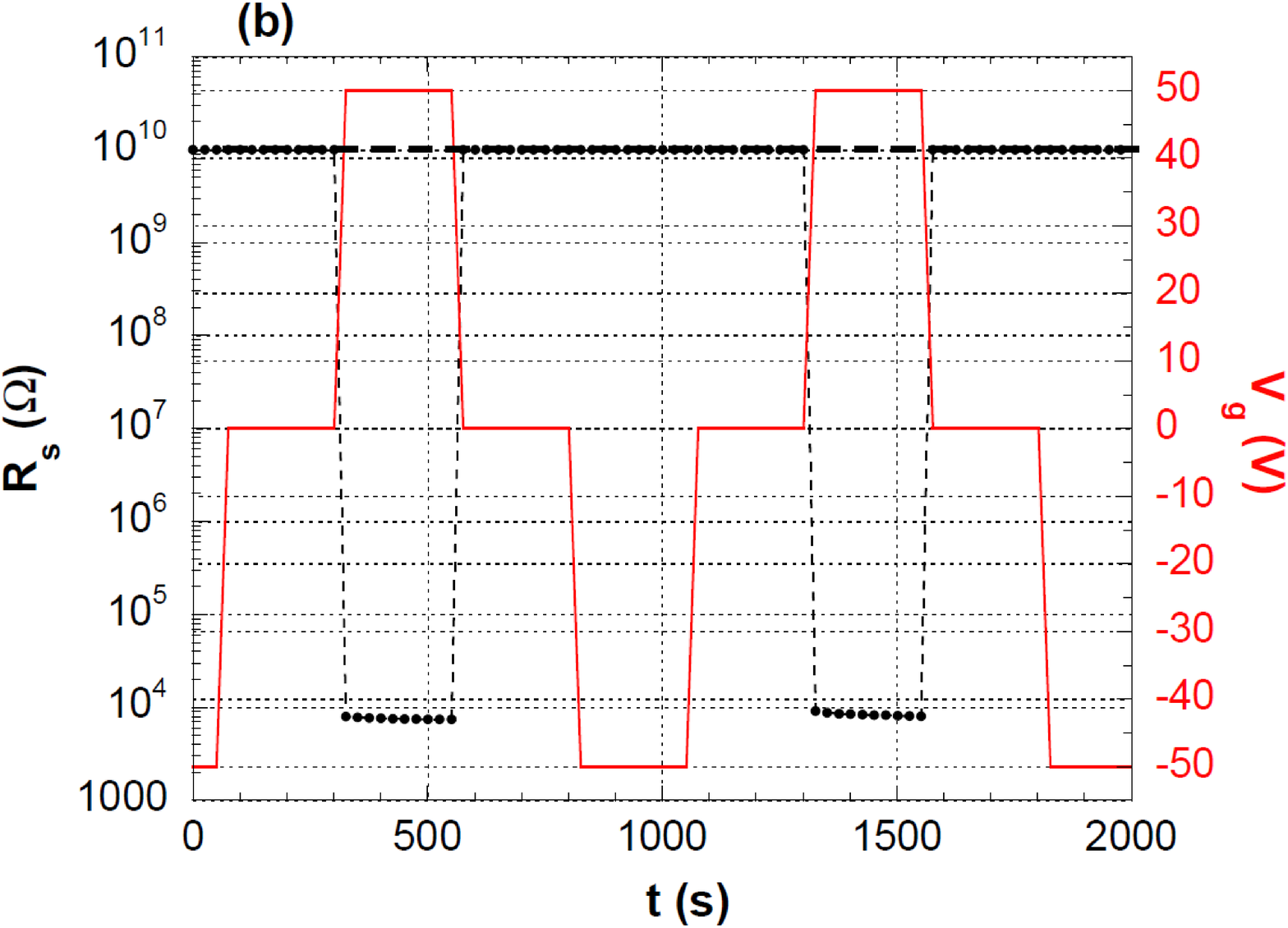}
\caption{(a) $R_s$ response to $V_g$ cycles at $4K$ for a low-$R_s$ sample $V_g$ tuned at room T ($R_{s300K} = 60k\Omega$, bias voltage 10mV). $R_s$ is immeasurably large (above $10G\Omega$) under $-30V$ and of only $7k\Omega$ under $30V$. (b) $R_s$ versus $V_g$ at $4K$ for the same low-$R_s$ sample $V_g$ tuned to $R_{s300K} = 120k\Omega$. $R_s$ is measurable only under $V_g = 50V$.}\label{Figure10}
\end{figure}

\begin{figure}[h]
% GVg4KMod3
\centering
\includegraphics[width=8cm]{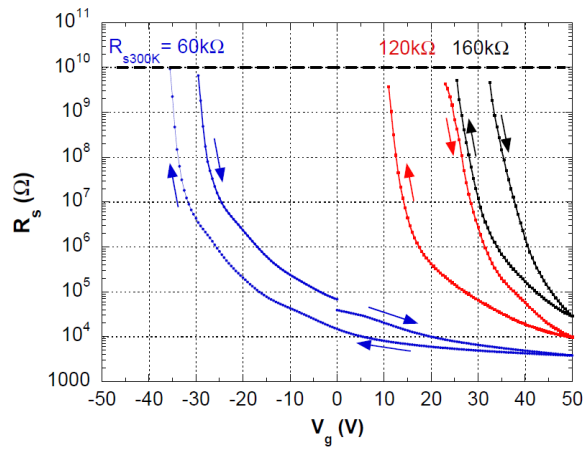}
\caption{$R_s$ versus $V_g$ at $4K$ for the sample already measured in Figure \ref{Figure10} and $V_g$ tuned at room temperature to three different $R_{s300K}$ values: $60k\Omega$, $120k\Omega$ and $160k\Omega$. $V_g$ was continuously swept from $0V$ to $50V$, then to $-50V$ and then back to $0V$ and the bias voltage was fixed to $100mV$. The curves are not plotted when $R_s$ exceeds the highest measurable value of our set-up (dotted line), which restrains the plotted curves to positive $V_g$ values for the two most resistive ones. }\label{Figure10b}
\end{figure}

But when $V_g$ tuned samples having larger $R_s$ values are measured, the electrical field effects change dramatically. The relative $R_s$ difference is about 50\% between $V_g = -30V$ and $30V$ at $4K$ when $R_{s300K} = 30k\Omega$ (non $V_g$ tuned sample) and it becomes as large as 6 orders of magnitude when $R_{s300K} = 60k\Omega$ following $V_g$ tuning (see Figure \ref{Figure10}). When $R_s$ is further increased, the resistance of the interface becomes unmeasurable under $V_g = 0V$ (transistor in the ``off'' state, $R_s > 10 G\Omega$) and is as low as $10k\Omega$ under $V_g = 50V$. Moreover and like in non $V_g$ tuned samples, the large $R_s$ modulations observed at $4K$ are accompanied by memory effects at $V_g = 0V$, which take the form of an hysteresis in the $R_s-V_g$ curves of Figure \ref{Figure10b}. Note that when $R_s$ increases, the $R_s-V_g$ values become also strongly bias dependent.

%with Low-$R_s$ samples tuned by $V_g$ at room T (see previous section) were used to adjust finely the low T properties of the AlOx/STO interface and to explore how the field effect evolves with $R_s$ at $4K$. We observed the following trends. First, the instantaneous variations of the resistance during $V_g$-cycles grows with $R_{s300K}$. The relative $R_s$ difference is about 50\% between $V_g = -30V$ and $30V$ at $4K$ when $R_{s300K} = 25k\Omega$ (non $V_g$-tuned sample) and is as large as 6 orders of magnitude when $R_{s300K} = 60k\Omega$ following $V_g$ tuning (see Figure \ref{Figure10}). The same trend is also observed when continuous $V_g$ scans are performed between $-50V$ and $50V$ for three different $R_{s300K}$ values (see Figure \ref{Figure10}). Second and like in non $V_g$-tuned samples, the large $R_s$ modulations observed at $4K$ are accompanied by memory effects at $V_g = 0V$, which take the form of an hysteresis in the $R_s-V_g$ curves of Figure \ref{Figure10}. Third, the $R_s-V_g$ values become strongly bias dependent when $R_s$ increases: for the largest $R_s$ values, the resistance at $V_g = 0V$ is measurable only in the non-linear regime.

Like at room temperature, the amplitude of the instantaneous field effect in non $V_g$ tuned low-$R_s$ samples is compatible with a standard charge injection process by the gate. According to our Hall effect measurements on a low-$R_s$ sample, we know that the surface charge density $n_s$ is almost constant with the temperature between $4K$ and $300K $\footnote{For a sample having $R_s = 100k\Omega$ at $300K$, $n_s$ was found to increase from $2\times10^{13}e/cm^2$ at $300K$ up to $2.4\times10^{13}e/cm^2$ at $80K$, before decreasing down to $1.6\times10^{13}e/cm^2$ at $4K$.}. But the dielectric constant of STO has a strong temperature dependence: $\epsilon_r$ is around 300 at $300K$, 2000 at $77K$, 20000 at $4K$ (under a small electric field) \cite{NevilleJAP72,MullerPRB79} and 10000 for $V_g = 50V$ (electrical field of $1kV/cm$) \cite{NevilleJAP72}. We thus expect the surface charge densities induced by the gate voltage $\Delta n_s(V_g)$ to be about ten times larger at $77K$ and 100 times larger at $4K$ than at $300K$ (at small enough $V_g$). It corresponds roughly to what is indeed measured for the low-$R_s$ sample of Figure \ref{Figure9} when $\Delta R_s/R_s$ is small: $\Delta R_s/R_s(50V) = 0.5\% $ at 300K, $\Delta R_s/R_s(50V) = 6\% $ at 77K and $\Delta R_s/R_s(5V) = 5\%$ at 4K. The fact that the $R_s$ modulation becomes larger in relative value in samples having increasing $R_s$ values can simply be explained by a decrease of their surface charge densities.

As for $V_g$ tuned high-$R_s$ samples, large resistance modulations at low temperatures have been reported in many studies on STO-based field effect devices in the back gate configuration \cite{NgaiPRB10,BellPRL09,CavigliaNature08}. Beyond the effect of a $n_s$ change, the mobility was also found to play a major role in the resistance modulations \cite{NgaiPRB10,BellPRL09,LiuAPLM15}. Under $V_g<0$, the electron gas is compressed closer to the interface and the mobility is lowered, while under $V_g>0$, the electron gas is extended towards the STO volume and the mobility is enhanced. The complete understanding of gating effects also requires the inclusion of the permittivity nonlinearities with the electric field \cite{BellPRL09}.

%Dire un peu plus : on commence, modulation faible. Puis modulation plus grande. On, off. si on augmente encore, off dans l'etat 0V. Obtenir ce qu'on veut comme effet de champ.

The origin of the memory effect and the hysteresis observed in Figures \ref{Figure9} and \ref{Figure10} remains unclear. Dielectric studies have shown that a ferroelectric state can be induced by a strong enough electrical field in STO crystals, due to the proximity of the ferroelectric transition \cite{MullerPRB79}. This ferroelectric state was usually found to be suppressed when T is increased beyond $50-100K$. Hysteresis and remnant polarization were measured at $1kV/cm$ in Ref. \cite{HembergerJPCM96,SaifiPRB70}, but also under smaller fields in LAO/STO interfaces \cite{CavigliaNature08,BellPRL09,RosslePRL13}. In Ref. \cite{RosslePRL13}, the resistance hysteresis was found to be related to a field induced polar order localized in the STO substrate about $1\mu m$ below the LAO/STO interface \cite{RosslePRL13}. Interestingly enough, small or no hysteresis was found at low T in field effect measurements in top-gate configuration \cite{HosodaAPL13}, indicating that the STO substrate plays the dominant role in the effect. In our case, a memory effect is visible at $4K$ already at $0.2kV/cm$ but not at $77K$ up to $1kV/cm$.

\section{Conclusion}

Our results show how the electrical properties of a metallic-like AlOx/STO interface can be tuned by the use of a back-gate voltage $V_g$. At room temperature, a slow increase of the resistance is observed under a negative $V_g$. This resistance increase can be large (many orders of magnitude) if $V_g$ is applied for a long enough time and the initial resistance value can be restored with the use of a positive $V_g$. The sample sheet resistance $R_s$ can thus be tuned on a controlled and reversible way to any value spanning from a metallic-like ($R_s=20k\Omega$, $dR/dT>0$ at $300K$) to an insulating-like behavior ($R_s>1M\Omega$, $dR/dT<0$ at $300K$). The activation energy of these slow resistance change time scales is about $0.7eV$ which strongly suggests that they are related to oxygen vacancy electro-migration. By pulling the oxygen vacancies in the volume of the STO crystal (the gate insulator material), the electrical field reduces the charge carriers density and mobility.
% At $4K$, the resistance response to $V_g$ changes is instead dominated by instantaneous effects related to the injection or the removing of charge carriers by $V_g$ at the interface. Irreversible resistance increase and memory effects are also observed under repeated $V_g$ cycles.

%We also demonstrate how the strong temperature dependence of this slow field effect around room T can be used to adjust the electrical properties of a single interface at low temperature.
If the interface resistance is $V_g$ tuned at room temperature and then rapidly cooled down below $250K$, a large set of stable interface electrical properties can be obtained. Such process can be used for example to increase the field effect amplitude at $4K$: fast resistance changes as large as six orders of magnitude can thus be obtained under $V_g$ cycles.
%We show that important parameters of a transistor, such as the amplitude of the resistance response to $V_g$ changes or the existence of an ``on'' or an ``off'' state at $V_g=0V$, can be adjusted in a single sample simply by applying at $300K$ a given $V_g$ for a given time.
We believe that this $V_g$ tuning technique at room temperature should also be applicable to the other 2D STO-based metallic systems where similar slow field effects have been reported. Beyond the adjustment of low temperature transistors parameters, it should allow the fine study of the metal-insulator transition or even the transition to the superconducting state as a function of the interface parameters (charge carriers density and mobility) in a single sample.

\vspace{1cm}
%\begin{thebibliography}{99}%

%\end{thebibliography}%

%\bibliography{SrTiO3IOPVg}

%\bibliography{SrTiO3IOPVg}

\end{document}